\documentclass[twocolumn,showpacs,showkeys]{revtex4}
\usepackage{graphicx}
\usepackage{bm}
\usepackage{color}
\usepackage{amsmath}
\usepackage{natbib}

\begin{document}

\title{On the interaction constant measurement of polarized fermions via sound wave spectra obtained from hydrodynamics with the pressure evolution equation}

\author{Pavel A. Andreev}
\email{andreevpa@physics.msu.ru}
\affiliation{Faculty of physics, Lomonosov Moscow State University, Moscow, Russian Federation, 119991.}

\date{\today}

\begin{abstract}
Usually, hydrodynamic equations are restricted by the continuity and Euler equations.
However,  the account of the higher moments of the distribution function gives better description of the kinetic properties.
Therefore, the pressure tensor evolution equation (PTEE) is derived for spin polarized degenerate fermions.
Moreover, it is found that the pressure tensor enters the interaction term generalizing the p-wave interaction in the Euler equation.
Hence, the PTEE allows to give a more accurate description of the interaction in Euler equation.
Next, the interaction is calculated for the PTEE.
The developed model is applied to the small amplitude bulk collective excitations in homogeneous and trapped fermions
in order to suggest the methods of experimental measurement of the interaction constant of polarized fermions.
It is demonstrated that the anisotropy in the momentum space revealing in the difference of the pressures in the anisotropy direction and the perpendicular directions leads to a method of detection of the interaction constant.
\end{abstract}

\pacs{03.75.Hh, 03.75.Kk, 67.85.Pq}
\keywords{degenerate fermions, hydrodynamics, pressure evolution equation, spinor quantum gas, separate spin evolution.}


\maketitle


Ultracold fermions is quantum fluids of neutral fermionic atoms
that are cooled into collective ground state
which is the subject  of much interest in recent years \cite{Drut PRL 18}.
Degenerate repulsive fermions,
spin-polarized fermions \cite{Minguzzi PRA 01},
weakly attractive fermions in BCS state \cite{Hannibal PRA 15},
BCS-BEC crossover \cite{Ries PRL 15, Boettcher PRL 16},
the unitary limit of highly interacting fermions
\cite{D.Lee PRB 06, Plantz PRA 19, Mukherjee PRL 19, Carcy PRL 19},
and boson-fermion mixtures \cite{Nakano PRB 16, Belemuk JPB 10}
show wide range of physical scenario in ultracold fermions.
Both the mean field short-range
\cite{Tylutki NJP 16, Antezza PRA 07},
and dipole-dipole long-range \cite{Babadi PRA 12}
interactions are studied in fermionic systems.
Fermions being in the single spin state are studied either \cite{Roth PRA 02, Roth PRA 01},
where $p$-wave interaction between fermions is considered due to the suppression of the s-wave scattering.

The dynamics of collective excitations can provide important information
about the nature and strength of particle interactions.
While the interaction define properties of nonlinear structures which can be found in degenerate fermions.

In spite of a large number of the theoretical and experimental researches,
there is no systematic analysis of the hydrodynamic (HD) model of degenerate fermions for the quantum gases.
We address this problem though the derivation of the HD equations from the microscopic many-body theory.
Being focused on the fermions in a single spin state (the regime of the full spin polarization)
we address the problem of interaction between fermions beyond the zero value s-wave scattering limit.

The majority of HD models deal with the continuity and Euler equations
(see for instance \cite{Kulkarni PRA 12},
where discussed different equations of state providing truncation of HD equation on the Euler equation stage
for the HD descriptions for some cold-atom systems in TABLE I).
However, the evolution of higher moments gives noticeable contribution \cite{Tokatly PRB 99, Tokatly PRB 00}.
Particularly,
if the sound waves are under consideration
it requires consideration of the pressure tensor evolution equation (PTEE).
The extended HD model suitable for the sound waves
in degenerate neutral fermions is addressed here.
The quantum HD method is used for the derivation \cite{Andreev PRA08, Andreev LP 19}.
The role of interaction in the spectrum of sound waves is described.

Small finite range of interaction reveals in a possibility of the force field expansion.
So, it appears as a series on the interparticle distance with the increasing order of derivative of the many-particle wave functions.
The first term appears in the first order by the interaction radius (the Gross-Pitaevskii approximation for the bosons).
Next term appears in the third order by the interaction radius.
The derivatives of the wave functions can be interpreted as the momentum operators.
Hence, the third order by the interaction radius gives result similar to the $p$-wave interaction \cite{Parker PRA 12}.
Corresponding energy density is presented in Ref. \cite{Parker PRA 12} (eq. 5),
a polynomial equation for the concentration of fermions is obtained in the Thomas-Fermi approximation (see \cite{Roth PRA 02} eq. 19).
However, presented here HD model with the PTEE appears
to be a generalization of the models presented in Refs. \cite{Roth PRA 02, Parker PRA 12}.

HD equations can be derived from the corresponding kinetic model.
However, the kinetics is not necessary intermediate step.
The HD equations can be directly derived from the quantum mechanics.
The derivation can be started from the definition of the concentration of particles \cite{Andreev PRA08}
\begin{equation}\label{aSRIffL concentration def} n(\textbf{r},t)=\int
dR\sum_{i}\delta(\textbf{r}-\textbf{r}_{i})\Psi^{*}(R,t)\Psi(R,t),\end{equation}
where $dR=\prod_{i=1}^{N}d\textbf{r}_{i}$ is the element of volume in $3N$ dimensional configurational space,
with $N$ is the number of particles,
and $\Psi(R,t)$ is the wave function of the system of particles.
Using the Schrodinger equation with the finite range interaction potential of neutral atoms,
it can be demonstrated that concentration (\ref{aSRIffL concentration def}) satisfies the continuity equation
$\partial_{t}n+\nabla\cdot \textbf{j}=0$,
where the definition of the current $\textbf{j}$ appears
(see eq. 3 of Ref. \cite{Andreev PRA08} for the definition).
Equation of the current evolution (Euler equation) can be calculated:
\begin{equation} \label{aSRIffL Euler eq 1 via j}
m\partial_{t}j^{\alpha}+m\partial_{\beta}\Pi^{\alpha\beta}
=-n\partial^{\alpha}V_{ext}+F^{\alpha}_{int}, \end{equation}
where
subindexes $\alpha$ and $\beta$ are used for the components of tensors and vectors in cartesian coordinates,
the Einstein rule on the summation on the repeating index is also assumed,
$m$ is the mass of particle,
$V_{ext}$ is the potential of the external field acting on particles,
$\Pi^{\alpha\beta}$ is the momentum flux containing the pressure tensor (see eq. 5 of Ref. \cite{Andreev PRA08} for the definition),
\begin{equation} \label{aSRIffL F alpha def via n2}
F^{\alpha}_{int}=-\int (\partial^{\alpha}U(\textbf{r}-\textbf{r}'))
n_{2}(\textbf{r},\textbf{r}',t)d\textbf{r}' \end{equation}
is the force field with the two-particle concentration:
\begin{equation} \label{aSRIffL n2 def} n_{2}(\textbf{r},\textbf{r}',t)=\int
dR\sum_{i,j\neq i}\delta(\textbf{r}-\textbf{r}_{i})\delta(\textbf{r}'-\textbf{r}_{j})\Psi^{*}(R,t)\Psi(R,t) ,\end{equation}
and $U(\textbf{r}-\textbf{r}')$ is the potential of the interaction between particles.

Described method allows to derive the HD equations for all fermions
or for fermions with chosen spin projection \cite{Andreev LPL 18, Andreev PRE 15}.
In both cases, we present the force field as a series on the small radius
since we explicitly consider the short radius of the interaction.

Considering the short-range interaction between the fermions with the same spin projection
find that the force field is the divergence of the quantum stress tensor $F^{\alpha}_{ss}=-\partial_{\beta}\sigma^{\alpha\beta}$,
where the quantum stress tensor is derived in the following form
\begin{equation}\label{aSRIffL sigma fer TOIR via pressure and I 0}
\sigma^{\alpha\beta}=\frac{m^{2}}{2\hbar^{2}}g_{2} I_{0}^{\alpha\beta\gamma\delta}np^{\gamma\delta}, \end{equation}
where $I_{0}^{\alpha\beta\gamma\delta}=\delta^{\alpha\beta}\delta^{\gamma\delta}+\delta^{\alpha\gamma}\delta^{\beta\delta}+\delta^{\alpha\delta}\delta^{\beta\gamma}$, $\delta^{\alpha\beta}$ is the Kronecker symbol,
$p^{\alpha\beta}$ is the pressure tensor,
and $g_{2}=\int r^{2}U(r)d\textbf{r}$,
with $U(r)=U_{\uparrow\uparrow}(r)=U_{\downarrow\downarrow}(r)$.
It appears in the third order by the interaction radius.
The quantum stress tensor equals to zero in the first order by the interaction radius
due to the antisymmetry of the wave function.
Equation (\ref{aSRIffL sigma fer TOIR via pressure and I 0})
gives an analog of $p$-wave scattering between fermions
of the same spin projection \cite{Roth PRA 02}, \cite{Roth PRA 01}. 
The resemblance is more clear if the equation of state for the pressure is used as the diagonal Fermi pressure.

Interaction between fermions with different spin projections is derived in the following form
\begin{equation} \label{aSRIffL F up to TOIR via n1 n2}
F^{\alpha}_{d}=-g_{\uparrow\downarrow}n_{(1)}\partial^{\alpha}n_{(2)}
-\frac{1}{2}g_{2,\uparrow\downarrow}n_{(1)}\partial^{\alpha}\triangle n_{(2)},\end{equation}
where $g_{\uparrow\downarrow}=\int U_{\uparrow\downarrow}(r)d\textbf{r}$,
and $g_{2,\uparrow\downarrow}=\int r^{2}U_{\uparrow\downarrow}(r)d\textbf{r}$.
Subindex $(2)$ refers to particles causing force
and subindex $(1)$ refers to particles moving under the action of this force.
Subindex $d$ describes interaction between different species.
An expression similar to (\ref{aSRIffL F up to TOIR via n1 n2}) is found for the boson-fermion interaction in Refs. \cite{Andreev PRA08}.
The last term in equation (\ref{aSRIffL F up to TOIR via n1 n2}) appears in the third order by the interaction radius.
At the interaction of bosons the third order by the interaction radius term contains higher derivatives \cite{Andreev PRA08, Braaten PRA 01}.
Hence, for the small amplitude plane wave find that
at repulsive interaction $g_{2,\uparrow\downarrow}>0$ the third order by the interaction radius term gives effective attraction
since the second spatial derivative leads to $-k^{2}$, where $k$ is the module of wave vector.
Opposite conclusion follows from equation (\ref{aSRIffL sigma fer TOIR via pressure and I 0}) for the interaction of fermions of the same spin projection.
Moreover, it does not related to the amplitude of perturbations.

Fermions with different spin projections are considered as two different species.
Therefore, there are two the continuity equations:
$\partial_{t}n_{s}+\nabla\cdot (n_{s}\textbf{v}_{s})=0$.
Next, the Euler equations have the following form:
$$mn_{s}(\partial_{t} +\textbf{v}_{s}\cdot\nabla)v^{\alpha}_{s} +n_{s}\partial^{\alpha}V_{ext}+\partial^{\beta}p_{s}^{\alpha\beta}=-g_{\uparrow\downarrow} n_{s}\partial^{\alpha}n_{s'}$$
\begin{equation}\label{aSRIffL Euler TOIR SSE spin s with set with p}
-\frac{g_{2,\uparrow\downarrow}}{2}n_{s}\partial^{\alpha}\triangle n_{s'} -g_{2}\frac{m^{2}}{2\hbar^{2}}I_{0}^{\alpha\beta\gamma\delta}\partial^{\beta}(n_{s}p^{\gamma\delta}_{s})
,\end{equation}
where $s=\uparrow,\downarrow$ and $s'\neq s$, so $s'$ presents the different spin projection.
The velocity field is introduced by the method described in Refs. \cite{Andreev LPL 18, Andreev PRE 15}.

At this step the two-particle concentrations are calculated in the weakly interacting limit.
The small radius of interaction is assumed and terms up to the third order by the interaction radius are included.
However, no assumption is made about the form of the pressure tensor $p^{\alpha\beta}$.
The distribution of degenerate fermions on quantum states in the momentum space requires an independent treatment of the collective variables
which are quantum mechanical average of the higher degree of the momentum operator
as the pressure tensor.
It differs from the bosons in the BEC state,
where there is no distribution of bosons on different quantum states.
So, the pair of HD equations present a suitable model which is equivalent to the Gross-Pitaevskii equation.

The PTEE for fermions with a chosen spin projection has the following form:
$$\partial_{t}p_{s}^{\alpha\beta} +v_{s}^{\gamma}\partial_{\gamma}p_{s}^{\alpha\beta} +p_{s}^{\alpha\gamma}\partial_{\gamma}v_{s}^{\beta} +p_{s}^{\beta\gamma}\partial_{\gamma}v_{s}^{\alpha}
+p_{s}^{\alpha\beta}\partial_{\gamma}v_{s}^{\gamma} $$
$$=-\frac{m}{8\hbar^{2}}g_{2} \{I_{0}^{\alpha\gamma\delta\mu}
[3 n_{s}^{2} v_{s}^{\beta}v_{s}^{\delta}\partial^{\gamma}v_{s}^{\mu}
+2n_{s}p_{s}^{\mu\delta}(\partial^{\gamma}v_{s}^{\beta}-\partial^{\beta}v_{s}^{\gamma})]$$
\begin{equation} \label{aSRIffL eq for p alpha beta II with set spin s}
+I_{0}^{\beta\gamma\delta\mu}
[3 n_{s}^{2} v_{s}^{\alpha}v_{s}^{\delta}\partial^{\gamma}v_{s}^{\mu}
+2n_{s}p_{s}^{\mu\delta}(\partial^{\gamma}v_{s}^{\alpha}-\partial^{\alpha}v_{s}^{\gamma})]\}.  \end{equation}
The external field (trapping potential) does not contribute in the PTEE.
The interaction term is presented in the Thomas-Fermi approximation,
where the higher spatial derivatives of the HD functions are neglected.

The derivation method is demonstrated in Ref. \cite{Andreev PRA08}.
Details of derivation of equations (\ref{aSRIffL sigma fer TOIR via pressure and I 0})-(\ref{aSRIffL eq for p alpha beta II with set spin s})
will be published elsewhere.
Equation (\ref{aSRIffL eq for p alpha beta II with set spin s}) describes evolution of six elements of symmetric pressure tensor.
Evolution of scalar pressure is considered in context of quantum gases in Ref. \cite{Csordas PRA 00},
the spectrum of small amplitude collective excitations of degenerate fermions in parabolic traps is described there.
Nonuniform scalar pressure contribution in the spectrum of collective modes in finite temperature bosons is discussed in Ref. \cite{Griffin PRL 97}.

The account of the PTEE and the interaction up to the third order by the interaction radius extends the mean-field model of degenerate fermions.
Such kind of extensions open up paths for discovering of new fundamental phenomena in quantum gases
similarly to new kinds of solitons \cite{Wang NJP 14, Andreev MPL B 12}
or formation of structures \cite{Pfau Nature 16, Baillie 16}.

In general case, a proper model of fermions would require
a further extension of the HD model including the quantum mechanical average of the third and forth degree of the momentum operator.
However, the structure of model depends on the considering problem.
If one needs to study the sound waves $\omega=u_{s}k$,
where $\omega$ is the frequency of wave, $k$ is the wave vector and $u_{s}$ is the speed of sound,
or waves with a gap in the spectrum $\omega^{2}=\omega^{2}_{0}+\beta k^{2}$,
where $\omega_{0}$ is the cut-off frequency and $\beta$  coefficient proportional to the square of the speed of sound,
and corresponding nonlinear excitations, the presented model gives a strong background in this regimes.
This conclusion does not follow directly from the derivation of the presented equations,
but it comes from the analysis of the extended model and its contribution to the spectrum of the bulk collective excitations
(see for instance \cite{Tokatly PRB 00}).

Consider the evolution of small amplitude bulk perturbations.
It shows correction of the sound wave expression due to the account of the pressure dynamics
rather use of the continuity and Euler equations.
Focusing on the full spin polarization we describe single species of fermions.

If the uniform medium is considered the following equilibrium conditions can be chosen:
nonzero partial concentrations $n_{0}$, zero partial velocity fields $\textbf{v}_{0}=0$, and nonzero partial pressures $p_{0}$ ($p_{0}^{\alpha\beta}=p_{0}\cdot\delta^{\alpha\beta}$).

Perturbation are considered in the form of the plane wave
in the three dimensional space $\delta n$, $\delta v^{\alpha}$, $\delta p^{\alpha\beta}$$\sim \exp(-\imath\omega t+\imath kx)$,
where the wave propagate parallel to the $x$ axis.
The longitudinal perturbations with the velocity field perturbations parallel to the direction of wave propagation
$\delta \textbf{v}\parallel \textbf{k}$ are considered.
In the chosen case $\textbf{k}= k \textbf{e}_{x}$.
Hence, this regime is related to the evolution of $x$-projection of the velocity field $\delta v_{x}$.

In this regime the HD equations simplifies to the following form.
Moreover, as it is mentioned above, the set of equations splits on two subsystems.
One of them describes the longitudinal perturbations:
$\delta n=k_{x}n_{0}\delta v_{x}/\omega$,
$$mn_{0}\omega\delta v_{x}-k_{x}\delta p^{xx}$$
\begin{equation} \label{aSRIffL velocity x lin}
-\Lambda k_{x}(n_{0}\delta p^{\alpha\alpha}+2n_{0}\delta p^{xx}+5p_{0}\delta n)/2n_{0}=0, \end{equation}
$\omega\delta p^{xx}-3p_{0}k_{x}\delta v_{x}=0$,
and
$\delta p^{yy}=\delta p^{zz}=k_{x}p_{0}\delta v_{x}/\omega$,
also appears from equation (\ref{aSRIffL eq for p alpha beta II with set spin s}),
where $\Lambda\equiv g_{2} mn_{0}/\hbar^{2}$ and $\delta p^{\alpha\alpha}=\delta p^{xx}+\delta p^{yy}+\delta p^{zz}$,
so all diagonal elements of pressure give contribution in $x$-projection of the velocity field via the interaction term.
The linear analysis provides an equation of state for perturbations of elements of the pressure tensor.
Equation of state appears to be different for different diagonal elements.
The element related to the direction of wave propagation $\delta p^{xx}=3p_{0}\delta n/n_{0}$ appears to be three times larger
then elements related to the perpendicular directions $\delta p^{yy}=\delta p^{zz}=p_{0}\delta n/n_{0}$.
It is shown that plane bulk excitation do not engage the interaction in the pressure evolution equation.

If there is full spin polarization find the single longitudinal acoustic wave with the following spectrum:
\begin{equation}\label{aSRIffL single species wave spectrum longitudinal}
\omega^{2}_{l}=\frac{3p_{0}}{mn_{0}}k^{2}\biggl(1+\frac{8}{3}\Lambda\biggr), \end{equation}
where equilibrium pressure can be used in the standard form for system of fermions located in the single spin state $p_{0}=(6\pi^{2})^{\frac{2}{3}}\hbar^{2}n_{0}^{\frac{5}{3}}/5m$.
Parameter $\Lambda$ contains dependencies on the species via the mass and interaction constant.
For a fixed species it has a dependence on the particle number $n_{0}$.
It appears as result of evolution of the concentration, $x$-projection of the velocity field,
and all diagonal elements of the pressure tensor.

Consider the noninteracting limit of equation (\ref{aSRIffL single species wave spectrum longitudinal}) using the presented expression for the pressure
and find $\omega^{2}_{l}=\frac{3}{5}(6\pi^{2})^{\frac{2}{3}}\hbar^{2}n_{0}^{\frac{2}{3}}k^{2}/m^{2}$.
It is fount that the frequency square is proportional to $(3/5)v_{Fe}^{2}$ in accordance with the kinetic theory of degenerate fermions.

In the traditional HDs,
where the PTEE is not considered obtain the following spectrum
$\omega^{2}_{l}=\frac{1}{m}\frac{\partial p_{0}}{\partial n_{0}}k^{2} +\frac{20}{3}\frac{p_{0}}{mn_{0}}\Lambda k^{2}$,
instead of equation (\ref{aSRIffL single species wave spectrum longitudinal}).
In the noninteracting limit it simplifies to $\omega^{2}_{l}=(6\pi^{2})^{\frac{2}{3}}\hbar^{2}n_{0}^{\frac{2}{3}}k^{2}/3m^{2}$,
where the Fermi pressure presented above is used as an equation of state for the equilibrium part of pressure.
Hence, the frequency square is proportional to $(1/3)v_{Fe}^{2}$.
Full expression becomes $\omega^{2}_{l}=\frac{5}{3}\frac{p_{0}}{mn_{0}}k^{2}(1+4\Lambda)$.
Moreover, there is a term proportional to the first order of the interaction constant
while equation (\ref{aSRIffL single species wave spectrum longitudinal}) contains the square of the interaction constant
which arises from the PTEE.

Isotropic equilibrium pressure considered above corresponds to simplest structure of the Fermi surface.
Majority of physical systems have nonspherical Fermi surface.
It provides more complex structure for the equilibrium pressure.
Consider a regime of diagonal pressure with $p_{0}^{xx}=p_{0}^{yy}\neq p_{0}^{zz}$,
which corresponds to anisotropy axis directed along $z$ axis.
We have two nontrivial directions: parallel to the anisotropy axis
(with $\delta p^{xx}=\delta p^{yy}=k_{z}p_{0}\delta v_{z}/\omega$
and $\delta p^{zz}=3k_{z}r_{0}\delta v_{z}/\omega$)
and perpendicular to the chosen direction
(with $\delta p^{xx}=3k_{x}p_{0}\delta v_{x}/\omega$
$\delta p^{yy}=k_{x}p_{0}\delta v_{x}/\omega$
and $\delta p^{zz}=k_{x}r_{0}\delta v_{x}/\omega$),
where $p_{0}^{xx}=p_{0}^{yy}\equiv p_{0}$, and $p_{0}^{zz}\equiv r_{0}$.

Propagation parallel to the anisotropy axis $\delta n$, $\delta v^{\alpha}$, $\delta p^{\alpha\beta}$$\sim e^{-\imath\omega t+\imath k_{z}z}$ gives linear dependence of frequency square $\omega^{2}$
on the dimensionless interaction constant for the longitudinal waves:
\begin{equation}\label{aSRIffL single species wave spectrum longitudinal anis par z}
\omega_{\parallel}^{2}=\frac{3k_{z}^{2}}{mn_{0}}\biggl(r_{0}(1+2\Lambda)+\frac{2}{3}p_{0}\Lambda\biggr).\end{equation}

Changing of the ratio $l=r_{0}/p_{0}$ we change relative contribution of the interaction in the wave spectrum.
This change allows to detect the contribution of interaction and extract the value of the interaction constant from the measurements of the small amplitude perturbation spectrum.

Next, consider propagation perpendicular to the anisotropy axis, where $\delta n$, $\delta v^{\alpha}$, $\delta p^{\alpha\beta}$$\sim e^{-\imath\omega t+\imath k_{x}x}$. It provides the following spectrum
\begin{equation}\label{aSRIffL single species wave spectrum longitudinal anis perp z}
\omega_{\perp}^{2}=\frac{3k_{x}^{2}}{mn_{0}}\biggl(p_{0}\biggl(1+\frac{7}{3}\Lambda\biggr)+\frac{1}{3}r_{0}\Lambda\biggr).\end{equation}
It shows different contribution of the interaction constant in coefficients in front of pressure elements $p_{0}$ and $r_{0}$.

Next adopt this formal result for the parabolic trap confined degenerate fermions in the regime of cigar shaped traps.
Start with the presentation of generalization of spectrum (\ref{aSRIffL single species wave spectrum longitudinal anis par z})
for the trapped fermions.
Consider the equilibrium condition.
At this point, let us notice
that the pressure evolution equation does not contain the confinement potential
and each term in the pressure evolution equation contains either velocity $\textbf{v}$
or the time derivative.
Therefore, it gives no contribution in the analysis of the equilibrium state.
As above, assume that the pressure tensor is diagonal in the equilibrium state
$p_{0}^{\alpha\beta}=\{p_{0}^{xx},p_{0}^{yy},p_{0}^{zz}\}$.
Moreover, assume
that the diagonal elements of the pressure tensor are functions of concentration like the Fermi pressure
$p_{0}^{xx}=p_{0}^{yy}=\alpha n^{5/3}$
and
$p_{0}^{zz}=\beta n^{5/3}$.
Then the interaction term in the Euler equation (\ref{aSRIffL Euler TOIR SSE spin s with set with p}) can be rewritten in the following form
$-g_{2}(4m^2/5\hbar^{2})\nu_{0}n\nabla n^{5/3}$,
where $\nu_{0}=\nu_{\perp}=4\alpha+\beta$ for the $x$- and $y$-projections of the Euler equation,
and $\nu_{0}=\nu_{\parallel}=2\alpha+3\beta$ for the $z$-projections of the Euler equation.
The last term of the Euler equation does not appear as the gradient of a potential even for the equilibrium state.
Hence, there is no Cauchy-Lagrangian integral.
The last term of the Euler equation can be written as a spatial derivative,
but it gives different "potential" for different projections of the Euler equation.

Spectrum (\ref{aSRIffL single species wave spectrum longitudinal anis par z}) is found for waves propagating parallel to the anisotropy direction
which is the $z$-direction.
Therefore, its generalization appears if the axis of the cigar shaped trap is directed parallel to the $z$-direction.
We consider regime of the strong confinement in the $x$- and $y$-directions and the weak confinement in the $z$-direction.
Hence, we have quasi-one dimensional dynamics in the $z$-direction.

If the external force in the strong confinement direction dominates over interaction contribution in the corresponding projections of the Euler equation
we drop the interaction in these projections of the Euler equation.
So, we have an effective nonlinear potential:
\begin{equation}\label{aSRIffL V eff} V_{eff}=\frac{m\omega_{\perp}^{2}}{2}\rho^{2}
+\frac{5}{2}\beta n^{2/3}+g_{2}\frac{4m^{2}}{5\hbar^{2}}(2\alpha+3\beta)n^{5/3},\end{equation}
where $\rho^{2}=x^{2}+y^{2}$.
Effective potential is obtained by replacement of coefficient $\nu_{\perp}$ by $\nu_{\parallel}$ in neglegible terms.

The effective potential allows to find the Cauchy-Lagrangian integral of the Euler equation for the motion with potential velocity field,
which is suitable for the sound wave perturbations:
$m\partial_{t}\varphi+m(\nabla\varphi)^{2}/2+V_{eff}=\mu$,
where $\mu$ is the chemical potential.
The Cauchy-Lagrangian integral together with the continuity equation leads to the non-linear Schrodinger equation for the macroscopic wave function of the concentration $n$ and the potential of the velocity field $\varphi$ ($\Phi=\sqrt{n}e^{\imath m\varphi/\hbar}$):
$\imath\hbar\partial_{t}\Phi=-\frac{\hbar^{2}}{2m}\triangle\Phi+V_{eff}\Phi$.
For the cigar-shaped trap the macroscopic wave function can be factorized:
$\Phi=\Phi_{0}(\rho)\Phi(z)$,
with $\Phi_{0}=e^{-\rho^{2}/2a^{2}}/a\sqrt{\pi}$, and $a=\sqrt{\hbar/m\omega}$.

\begin{figure}
\includegraphics[width=8cm,angle=0]{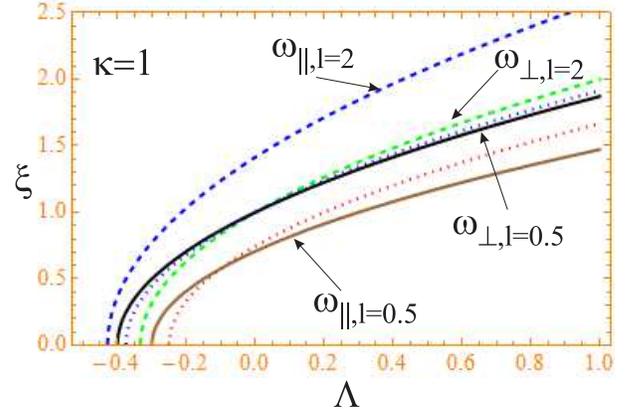}
\caption{\label{aSRIffL 01} Frequency of sound wave propagating parallel or perpendicular to the anisotropy axis of the Fermi surface
as the function of the dimensionless interaction constant $\Lambda$ is demonstrated.
The dimensionless wave vector is introduced via the Fermi wave vector $k_{F}=(6\pi n_{0})^{1/3}$:
$\kappa=k_{i}/k_{F}$,
where $i=x$ or $z$.
Corresponding dimensionless frequency is $\xi=\omega/v_{p}k_{F}$,
where $v_{p}=\sqrt{3p_{0}/mn_{0}}$.
Red (the lower dotted) line describes regime with no pressure evolution;
the lower dashed green (perpendicular to anisotropy axis of the Fermi surface)
and the upper dashed blue (parallel to the anisotropy axis) lines present the regime of anisotropic Fermi surface for $l=2$;
the upper continuous black (perpendicular to anisotropy axis)
and the lower continuous brown (parallel to the anisotropy axis) lines show the regime of anisotropic Fermi surface for $l=0.5$;
blue (the upper dotted) line corresponds to regime of isotropic Fermi surface.}
\end{figure}

Quasi-one-dimensional non-linear Schrodinger equation reads
$$\imath\hbar\partial_{t}\Phi(z)=-\frac{\hbar^{2}}{2m}\frac{d^{2}\Phi(z)}{dz^{2}}+\beta' \mid\Phi(z)\mid^{4/3}\Phi(z)$$
\begin{equation}\label{aSRIffL NLSE}
+g_{2}\gamma_{1}\frac{4m^{2}}{5\hbar^{2}}\nu_{\parallel}\mid\Phi(z)\mid^{10/3}\Phi(z),\end{equation}
where
$\beta'=(3/2)\sqrt[3]{\pi a^{2}}\beta$,
and $\gamma_{1}=3/(5a\sqrt{\pi}\sqrt[3]{a\sqrt{\pi}})$.

Equation (\ref{aSRIffL NLSE}) allows to find equilibrium quasi-one-dimensional macroscopic wave function.
If we need to consider quasi-one-dimensional dynamics
we should go back to hydrodynamic equations knowing transverse behaviour of functions.

Quasi-one-dimensional continuity equation is obtained in the obvious form
$\partial_{t}\tilde{n}(z)+\partial_{z}(\tilde{n}(z)v_{z}(z))=0$.
Extraction of the transverse equilibrium parts of fermions and further integration over the transverse coordinates
leads to the following quasi-one-dimensional Euler equation:
\begin{equation}\label{aSRIffL Euler TOIR SSE spin s with set with p QOD}
\tilde{n}(\partial_{t}+v_{z}\partial_{z})v_{z} +\frac{\gamma_{1}}{m}\partial_{z}\tilde{p}^{zz}
=\frac{3mg_{2}}{16\hbar^{2}}\gamma_{2}^{4}\partial_{z}[\tilde{n}(\tilde{p}^{\gamma\gamma}+2\tilde{p}^{zz})],\end{equation}
where $\gamma_{2}\equiv 1/\sqrt[3]{a}\sqrt[6]{\pi}$, and $\tilde{n}=\tilde{n}(z)$.
Here and below, $\tilde{p}^{\alpha\beta}=\tilde{p}^{\alpha\beta}(z)$ is a part of pressure
(the equilibrium transverse dependence is extracted from the full pressure
$p^{\alpha\beta}(\textbf{r})=n_{0}(\rho)^{5/3}\tilde{p}^{\alpha\beta}(z)$).
$z$-projection of the velocity field is the single nonzero projection which is a function of one projection of coordinate
$v_{z}=v_{z}(z)$.

After integration of the PTEE on direction perpendicular to the trap axis direction ($x$- and $y$-directions)
obtain it in the following form:
$$\partial_{t}\tilde{p}^{\alpha\beta}+v_{z}\partial_{z}\tilde{p}^{\alpha\beta}
+(\tilde{p}^{\alpha z} \delta^{\beta z}+\tilde{p}^{\beta z}\delta^{\alpha z}+\tilde{p}^{\alpha\beta})\partial_{z}v_{z}$$
\begin{equation}\label{aSRIffL QOD pressure evol}  =-\frac{15m}{8\hbar^{2}}g_{2}\gamma_{2}^{2}\delta^{\alpha z}\delta^{\beta z}\tilde{n}^{2}v_{z}^{2}\partial_{z}v_{z}.\end{equation}
The right-hand side of equation (\ref{aSRIffL QOD pressure evol}) is nonlinear.
Hence, it gives no contribution in the spectrum.

Equations (\ref{aSRIffL Euler TOIR SSE spin s with set with p QOD}) and (\ref{aSRIffL QOD pressure evol}) give
the following generalization of spectrum (\ref{aSRIffL single species wave spectrum longitudinal anis par z}):
\begin{equation}\label{aSRIffL single species wave spectrum longitudinal anis par z trap}
\omega_{\parallel}^{2}=\frac{3k_{z}^{2}\gamma_{1}}{m\tilde{n}_{0}}\biggl(\tilde{r}_{0}+\frac{10}{3}\Lambda \biggl(\tilde{r}_{0}+\frac{1}{3}\tilde{p}_{0}\biggr)\biggr),\end{equation}
where $\gamma_{2}^{4}/\gamma_{1}=5/3$.
Additional coefficient in front of terms describing interaction appears to different dependence of the interaction terms and pressure term on equilibrium concentration.

\begin{figure}
\includegraphics[width=8cm,angle=0]{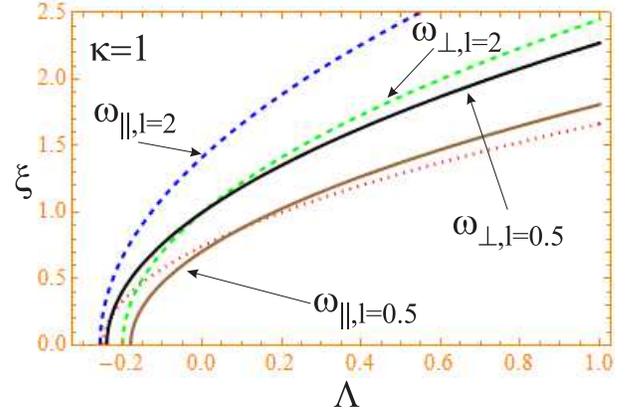}
\caption{\label{aSRIffL 02} Frequency of sound wave propagating parallel or perpendicular to the anisotropy axis of the Fermi surface
as the function of $\Lambda$ is demonstrated.
The trap frequency is chosen in a way that the sound speed of trapped gas
$\tilde{v}_{p}=\sqrt{3\tilde{p}_{0}\gamma_{1}/mn_{0}}$ to fit to the sound speed of untrapped gas $v_{p}$.
The dotted red line describes regime with no pressure evolution.
Notations of other curves is the same as Fig. \ref{aSRIffL 01}.}
\end{figure}

\begin{figure}
\includegraphics[width=8cm,angle=0]{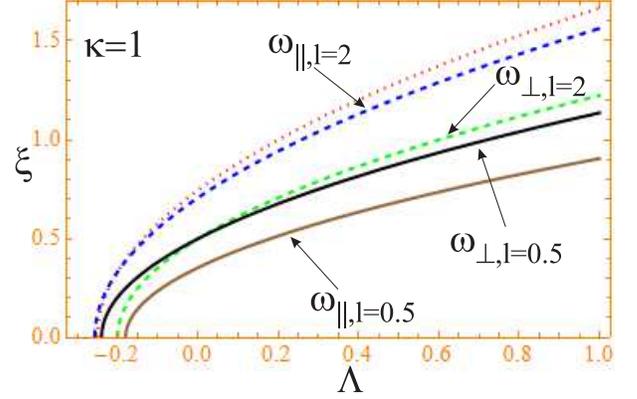}
\caption{\label{aSRIffL 03} Frequency of sound wave propagating parallel or perpendicular to the anisotropy axis of the Fermi surface
as the function of $\Lambda$ is demonstrated.
The figure is made for $\tilde{v}_{p}=v_{p}/2$.
Notations of curves is the same as Figs. \ref{aSRIffL 01}, \ref{aSRIffL 02}.}
\end{figure}

Similarly, we can introduce a cigar-shaped trap with the axis perpendicular to the anisotropy direction of the Fermi surface.
It corresponds to replacement of coordinate $z$ by $x$,
coefficient $\beta'$ by $\alpha'=(3/2)\sqrt[3]{\pi a^{2}}\alpha$,
and coefficient $\nu_{\parallel}$ by $\nu_{\perp}$ in equation (\ref{aSRIffL NLSE}) for the equilibrium state.
Coordinate $z$ should be replaced by coordinate $x$ in all elements of equations (\ref{aSRIffL V eff})-(\ref{aSRIffL NLSE}),
while general form of dynamical equations does not change.
Going through the steps similar to described above find a generalization of spectrum
(\ref{aSRIffL single species wave spectrum longitudinal anis perp z}) for the trapped fermions.


After all obtain an analog of spectrum (\ref{aSRIffL single species wave spectrum longitudinal anis perp z}) for quasi-one-dimensional trap:
\begin{equation}\label{aSRIffL single species wave spectrum longitudinal anis perp z trap}
\omega_{\perp}^{2}=\frac{3k_{x}^{2}\gamma_{1}}{m\tilde{n}_{0}}\biggl(\tilde{p}_{0}+\frac{5}{9}\Lambda(7 \tilde{p}_{0}+\tilde{r}_{0})\biggr),\end{equation}

Three numerical regimes are presented in Figs. \ref{aSRIffL 01}, \ref{aSRIffL 02}, \ref{aSRIffL 03}.
Each regime is made for two forms of the Fermi surfaces:
the cigar shaped Fermi surface elongated in $z$-direction,
and the pancake shaped Fermi surface flattened in $z$-direction.
The account of the pressure evolution at the isotropic Fermi surface (the Fermi sphere)
increases the frequency in compare with the traditional hydrodynamics (see Fig. \ref{aSRIffL 01}).
Propagation perpendicular to the anisotropy axis of the Fermi surface leads
to a small nonmonotonic deviation from the isotropic regime.
Transition to the trapped regime leads to shift of curves towards larger interaction constants and increase of frequency (see Fig. \ref{aSRIffL 02}).
An increase of the trap strength reduces the frequency with no shift of curves (see Fig. \ref{aSRIffL 03}).


Next summarize the obtained results.
A HD model of degenerate partially spin polarized neutral fermions has been obtained for a regime
where there are the concentration, velocity field and pressure tensor evolution equations are involved.
Moreover, the short-range interaction between fermions with the same spin projection has been calculated up to the third order by the interaction radius.
It is presented in the Euler and pressure evolution equations.
The model has been adopted for the parabolic trap.
It opens a possibility for study of linear and nonlinear collective excitations and role of fermion-fermion interaction in their formation.
It appears to be a simplified model in compare with kinetic equations for analytical and numerical analysis.

The spectrum of bulk collective excitations has been considered in terms of obtained model
for the uniform medium and cigar shaped traps in the regime of anisotropic Fermi surface
to demonstrate the role of interaction.


\begin{thebibliography}{17}




\bibitem{Drut PRL 18} J. E. Drut, J. R. McKenney, W. S. Daza, C. L. Lin, and C. R. Ordonez,
Phys. Rev. Lett. \textbf{120}, 243002 (2018).

\bibitem{Minguzzi PRA 01} A. Minguzzi, P. Vignolo, M. L. Chiofalo, and M. P. Tosi,
Phys. Rev. A \textbf{64}, 033605 (2001).


\bibitem{Hannibal PRA 15} S. Hannibal, P. Kettmann, M. D. Croitoru, V. M. Axt, and T. Kuhn,
Phys. Rev. A \textbf{98}, 053605 (2018).

\bibitem{Ries PRL 15} M. G. Ries, A. N. Wenz, G. Zurn, L. Bayha, I. Boettcher, D. Kedar, P. A. Murthy, M. Neidig, T. Lompe, and S. Jochim,
Phys. Rev. Lett. \textbf{114}, 230401 (2015).


\bibitem{Boettcher PRL 16} I. Boettcher, L. Bayha, D. Kedar, P. A. Murthy, M. Neidig, M. G. Ries, A. N. Wenz, G. Zurn, S. Jochim, and T. Enss,
Phys. Rev. Lett. \textbf{116}, 045303 (2016).
%

\bibitem{D.Lee PRB 06} D. Lee,
Phys. Rev. B \textbf{73}, 115112 (2006).
%

\bibitem{Plantz PRA 19} N. W. M. Plantz and H. T. C. Stoof,
Phys. Rev. A \textbf{99}, 013606 (2019).
%

\bibitem{Mukherjee PRL 19} B. Mukherjee, P. B. Patel, Z. Yan, R. J. Fletcher, J. Struck, and M. W. Zwierlein,
Phys. Rev. Lett. \textbf{122}, 203402 (2019).
%

\bibitem{Carcy PRL 19} C. Carcy, S. Hoinka, M. G. Lingham, P. Dyke, C. C. N. Kuhn, H. Hu, and C. J. Vale,
Phys. Rev. Lett. \textbf{122}, 203401 (2019).
%


\bibitem{Nakano PRB 16} E. Nakano, H. Yabu,
Phys. Rev. B \textbf{93}, 205144 (2016).

\bibitem{Belemuk JPB 10} A. M. Belemuk and V. N. Ryzhov,
J. Phys. B: At. Mol. Opt. Phys. \textbf{43}, 225301 (2010).

\bibitem{Tylutki NJP 16} M. Tylutki, A. Recati, F. Dalfovo and S. Stringari,
New J. Phys. \textbf{18}, 053014 (2016).
%

\bibitem{Antezza PRA 07} M. Antezza, F. Dalfovo, L. P. Pitaevskii and S. Stringari,
Phys. Rev. A \textbf{76}, 043610 (2007).

\bibitem{Babadi PRA 12} M. Babadi and E. Demler,
Phys. Rev. A \textbf{86}, 063638 (2012).
%


\bibitem{Roth PRA 02} R. Roth,
Phys. Rev. A \textbf{66}, 013614 (2002).


\bibitem{Roth PRA 01} R. Roth and H. Feldmeier,
Phys. Rev. A \textbf{64}, 043603 (2001).
%


\bibitem{Kulkarni PRA 12} M. Kulkarni, and A. G. Abanov,
Phys. Rev. A \textbf{86}, 033614 (2012).
%


\bibitem{Tokatly PRB 99} I. Tokatly, O. Pankratov,
Phys. Rev. B \textbf{60}, 15550 (1999).
%


\bibitem{Tokatly PRB 00} I. V. Tokatly, O. Pankratov,
Phys. Rev. B \textbf{62}, 2759 (2000).
%


\bibitem{Andreev PRA08} P. A. Andreev, L. S. Kuz'menkov,
Phys. Rev. A \textbf{78}, 053624 (2008).

\bibitem{Andreev LP 19} P. A. Andreev,
Laser Phys. \textbf{29}, 035502 (2019).



\bibitem{Parker PRA 12} N. G. Parker, D. A. Smith,
Phys. Rev. A \textbf{85}, 013604 (2012).
%


\bibitem{Andreev LPL 18} P. A.  Andreev,
Laser Phys. Lett. \textbf{15}, 105501 (2018).


\bibitem{Andreev PRE 15} P. A. Andreev,
Phys. Rev. E \textbf{91}, 033111 (2015).
%

\bibitem{Braaten PRA 01} E. Braaten, H.-W. Hammer, and S. Hermans,
Phys. Rev. A \textbf{63}, 063609 (2001).


\bibitem{Csordas PRA 00} A. Csordas, R. Graham,
Phys. Rev. A \textbf{63}, 013606 (2000).

\bibitem{Griffin PRL 97} A. Griffin, W.-C. Wu, and S. Stringari,
Phys. Rev. Lett. \textbf{78}, 1838 (1997).

%
\bibitem{Wang NJP 14} Z. Wang, M. Cherkasskii, B. A Kalinikos, L. D. Carr, M. Wu,
New J. Physics \textbf{16}, 053048 (2014).


\bibitem{Andreev MPL B 12} P. A. Andreev, L. S. Kuzmenkov,
Mod. Phys. Lett. B \textbf{26}, 1250152 (2012).


%
\bibitem{Pfau Nature 16} H. Kadau, M. Schmitt,	M. Wenzel, C. Wink, T. Maier, I. Ferrier-Barbut, T. Pfau,
Nature \textbf{530}, 194 (2016).


%
\bibitem{Baillie 16} D. Baillie, R. M. Wilson, R. N. Bisset, and P. B. Blakie,
Phys. Rev. A \textbf{94}, 021602(R) (2016).






%
\end{thebibliography}
\end{document}